\documentclass[fleqn,10pt]{SelfArx} 

\definecolor{color1}{RGB}{0,0,90} 
\definecolor{color2}{RGB}{0,20,20} 

\usepackage{hyperref} 
\hypersetup{hidelinks,colorlinks,breaklinks=true,urlcolor=color2,citecolor=color1,linkcolor=color1,bookmarksopen=false,pdftitle={Title},pdfauthor={Author},urlcolor=blue}

\usepackage{graphicx}
\usepackage[square]{natbib}
\usepackage{amsmath}
\usepackage{multirow}
\usepackage{subcaption}
\usepackage{pdflscape}
\usepackage{epstopdf}
\usepackage[final]{pdfpages}

\usepackage{graphicx}
\usepackage{natbib}
\usepackage{amsmath}
\usepackage{amsfonts}
\usepackage{multirow}
\usepackage{tikz}
\usetikzlibrary{calc}

\usepackage{mathtools,amssymb,lipsum}

\usepackage{cuted}
\usepackage{boondox-calo}

\usepackage{calligra}

\newcommand{\bs}[1]{\boldsymbol{#1}}

\newcommand{\mc}[1]{\mathcal{#1}}

\JournalInfo{Published in Journal of Magnetism and Magnetic Materials, Vol. 657, 174585, 2026} 
\Archive{\href{http://dx.doi.org/10.1016/j.jmmm.2026.174585}{DOI: 10.1016/j.jmmm.2026.174585}} 

\PaperTitle{Coils in thermomagnetic harvesters - a comparative study} 

\Authors{Aske Chris Nilsson, 
Andrea Roberto Insinga, Salvatore De Angelis, Georgios Potsios and Rasmus Bjørk} 
\affiliation{\textit{Department of Energy Conversion and Storage, Technical University of Denmark - DTU, DK-2800 Kgs. Lyngby, Denmark}} 
\affiliation{\textbf{Corresponding author}: achni@dtu.dk} 

\Keywords{} 
\newcommand{\keywordname}{Keywords} 

\Abstract{Thermomagnetic generators (TMGs) are devices that convert waste heat to electricity through a change in magnetization of a solid material. This causes a changing flux through a coil, which induces an electromotive force per Faraday's law. However, the influence of the coil on the performance of the TMG has not been investigated and existing TMG prototypes merely utilize some coil, not the optimal coil for a given device. In this work we present an analytical and numerical model of a TMG that calculates power by explicitly coupling the TMGs magnetic and electric circuits and use this to analyze the influence of the coil on the TMG performance. We show that TMG power has a linear dependence on coil volume, independent of the specific combination of wire radius and coil turns. The model is validated with experimental data, and finally used to study prototype TMGs presented in literature, where we show that the power of these literature TMGs can be increased by a factor of 10-400 times, had larger coils been used in the prototypes.  }

\begin{document}

\flushbottom 

\maketitle 


\thispagestyle{empty} 
\section{Introduction}
\noindent As global energy demand rises, improving energy efficiency in industry, transportation, and power generation is crucial for a future sustainable development. An estimated 50\% of the global energy input is lost as waste heat, with nearly 60\% being low-grade ($<$100°C)\cite{Firth2019QuantificationEffects}, highlighting the need for both an efficient and scalable energy recovery method. However, energy conversion of low-grade waste heat presents significant challenges that limits its practical application and implementation due to the low exergy and Carnot efficiency associated with the temperatures near room temperature and temperature differentials\cite{Hur2023Low-gradeHarvesting}. 
Various technologies have been explored to address this challenge, including thermoelectric generators (TEGs)\cite{qian2024thermodynamics}, organic Rankine cycle (ORC)\cite{Wang2020Multi-objectiveRecovery}, and pyroelectric (PE) systems\cite{pandya2019new}. However, these technologies  are often limited by low conversion efficiencies, low power output and high material or device cost. 
A novel alternative technology is thermomagnetic generators (TMGs), originally envisioned by N. Tesla and T. A. Edison in the late 19th century \cite{N.Tesla1889Thermo-MagneticMotor, T.A.Edison1888PyromagneticMotor}, which has regained scientific interest due to recent advancements in magnetocaloric materials (MCM) and magnetic refrigeration at room temperature \cite{deOliveira2026ExperimentalContribution,franco2018magnetocaloric, Chen2022EvaluationHeat}. In its simplest form, a TMG can be understood as a thermal engine that converts heat input into magnetic work\cite{solomon1991design}. This magnetic work produced stems from the characteristic ferromagnetic phase transition of the MCMs. When the temperature of the material is increased above its characteristic Curie temperature, the material undergoes a phase transition from ferromagnetic to paramagnetic behavior. Only when the material is in its ferromagnetic state is it significantly attracted to a permanent magnet (PM), a phenomenon that can be exploited to generate a mechanical motion between a MCM and a PM in an actuating or rotating (motor) setup\cite{Kaneko2021DesignResults,Silva2025NovelResults,Ahmim2021Self-oscillationGenerator}. This motion can then be converted into electrical energy by conventional means. Alternately, induction can be used to directly exploit the change in the magnetic properties of the MCM to induce an electromotive force (emf) in an electric circuit\cite{Hur2023Low-gradeHarvesting}. This approach has the clear advantage that the resulting energy harvesting device does not require moving parts. While it would in principle be possible to only use the spontaneous magnetization occurring when the MCM becomes ferromagnetic, this would be a relatively weak effect. Instead, it is more convenient to rely on its drastic increase in magnetic susceptibility, and use a permanent magnet as a fixed field source. In order to minimize magnetic leakage, this is more effectively done in a closed-magnetic circuit configuration, i.e. an iron core such as those found in transformers. The focus of this study will be a device working on this principle.

In such a TMG one or several coils are wound around the magnetic circuit. In such a coil, which is externally connected to a load resistance, an electric current will flow according to Ohm’s law, with the induced emf scaled with the number of coil turns $N$ as described by Faraday’s law,
\begin{equation}
    \mc{E} = - N \frac{d\Phi}{dt}
    \label{faradays_law}
\end{equation}
where $\mc{E}$ is the emf and $\Phi$ is the magnetic flux in the magnetic circuit. Early theoretical calculations by L. Brillouin and H. P. Iskenderian suggested that TMGs could achieve Carnot efficiencies of up to $\sim{}55\%$\cite{Brillouin1948ThermomagneticGenerator}, and with more recent numerical modeling of the thermodynamic cycle predicting Carnot efficiencies exceeding $25\%$ at maximum power output\cite{Almanza2017NumericalCycle, Almanza2017First-Conversion}. Despite these promising theoretical predictions, experimental realizations reported in the literature have so far achieved only milliwatt-scale power outputs with Carnot efficiencies of $1.75\cdot10^{-3}\%$, $6\%$ and $14\%-18\%$ \cite{Waske2019EnergyTopology,Bahl2024DesignHarvester, Liu2023High-performanceSwitch}. The reason for these low reported power output and efficiencies must be understood in terms of the components of the TMG - the coil, permanent magnet and MCM material, but also crucially in their interplay. Merely having highly optimized individual components will not make the TMG as a whole have a high performance, but instead the entire set of device parameters should be considered simultaneously\cite{Liu2023SignificantRecovery}.

Previously, numerical models of TMGs have been used for optimizing the power output of an already realized TMG or in the early stages of designing a new setup, with the parameters optimized being the frequency, the size of the magnetic components, the temperature differences between the reservoirs, the mean temperature and the load resistance \cite{Jiang2022NumericalRecovery, Liu2023High-performanceSwitch,Dzekan2021CanGenerator, Bahl2024DesignHarvester}. 
The modeling in Refs. \cite{Dzekan2021CanGenerator,Bahl2024DesignHarvester} considered a magnetic flux analysis for an experimental TMG system, to optimize power and study the magnetic flux leakage, but no electrical coupling was present. 
In general, the theoretical modeling of the TMG has not considered the coupling of the magnetic and electric systems except from the early work by Brillouin and Iskenderian \cite{Brillouin1948ThermomagneticGenerator} and the more recently presented work by Jiang et. al. \cite{Jiang2022NumericalRecovery}. In these two studies, the principles of flux conservation were used to estimate the magnetic flux in the coils, which was then used in Faraday's law and the equations of the electric circuit to analytically estimate the induced current. However, the two studies do not analyze and solve the magnetic circuit in order to obtain these fluxes, which limits the influence that the shape and size of the magnetic components i.e. the magnetic reluctances have on the magnetic flux in the circuit. Furthermore, the maximum power transfer theorem\cite{Zeng2021MaximumTransfer} was not considered in Jiang et al. when deriving the analytic equation for maximum power output. If considered, it is clear that implementing a load capacitor in the electric circuit, matching the self-inductance of the coils, would result in larger power and higher efficiency, as discussed subsequently. 

In order to investigate the influence of the individual TMG components, and crucially their interplay, the aim of this work is to establish a general theoretical approach to explicitly couple the magnetic and electric circuits of a TMG, including previously neglected effects such as magnetic reluctance, load capacitor, magnetic saturation and a full coupling between the two circuits compared to the partial coupling presented in Refs. \cite{Brillouin1948ThermomagneticGenerator,Jiang2022NumericalRecovery}. To achieve this, we first analyze each circuit independently and then derive the governing equations that links them. The relevant physical mechanisms can be modeled with different dimensionality: a three-dimensional model being the most realistic but most computationally demanding case, which could be realized by, e.g., finite element methods. However, since one of our main goals is to develop a model that provides insight and even intuition into the effect of all the relevant design parameters, we adopt a zero-dimensional (lumped element) framework, i.e. a circuit model. The resulting set of coupled governing equations are thus ordinary differential equations that can be formulated in the time domain or the frequency domain. The equations can then be solved analytically under certain simplifying assumptions, or numerically integrated by standard numerical methods. The analytical solutions, although only valid in the limit of small susceptibility changes, are extremely valuable in terms of clarity of the results and the design criteria that they provide. 

As a case study, we demonstrate that the coupled models provide key insights into the role of coil design in optimizing performance of a TMG. We apply our model to evaluate existing experimental prototypes reported in the literature, assessing their coil volume choices in light of the insights provided by our framework.

While most of our study is centered around the circuit model just discussed, we also coupled the lumped-element 0D description of the electric circuit with a spatially-resolved model of the magnetic circuit in 2D. This hybrid 0D/2D model, which we implemented through finite element methods in the COMSOL Multiphysics environment, can also account for magnetic flux losses, magnetic leakage, and other geometric effects of the various system components. 

\begin{figure}[!ht]
    \centering
    \includegraphics[width=0.9\linewidth]{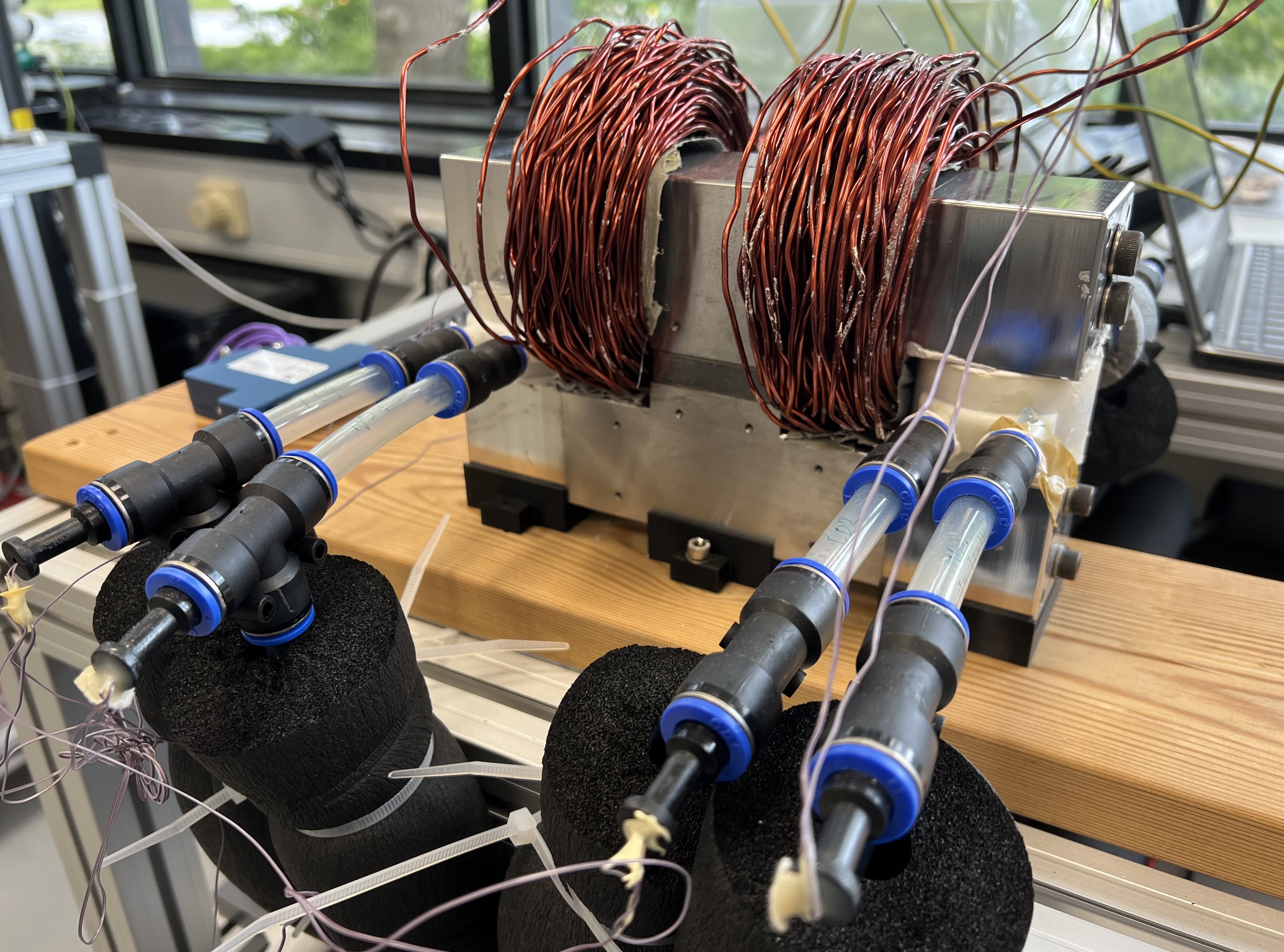}
    \caption{Static TMG prototype realized in previous work Ref.\cite{Bahl2024DesignHarvester}. The TMG has a figure-of-eight geometry and is assembled forming two loops, where a PM is placed in the middle and two MCM beds of packed gadolinium spheres on each loop side. Soft iron yoke is connecting the components and closing the two loops. Two copper coils are wound around the iron yoke, one on each loop, and connected with a load resistor. A set of solenoid valves and two pumps create an alternating hot and cold fluid stream, which is guided through the MCM beds and between two Julabo thermal baths by the plastic tubes.}
    \label{Experimental_setup}
\end{figure}

\section{Coupled model of a TMG}
To understand the interplay of the different components in a TMG, we begin by analyzing the magnetic and electric circuits independently. From this, we derive the governing equations for each circuit and show how they can be coupled to describe the induced current in the coils. \newline

\begin{figure}[!ht]
    \centering
    \includegraphics[width=0.99\linewidth]{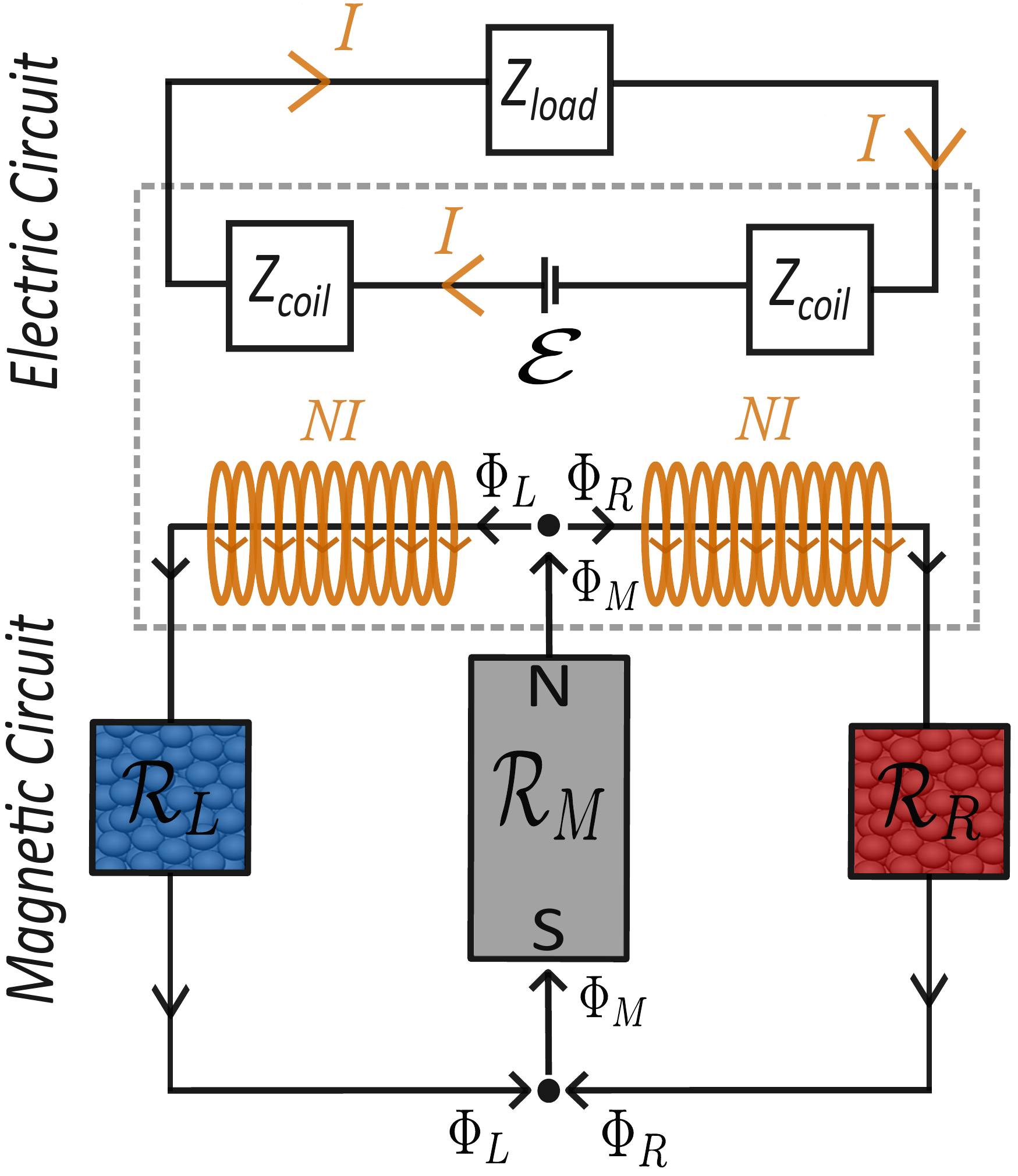}
     \caption{The coupled magnetic and electric circuit. Bottom: The magnetic circuit is composed of 1 permanent magnet in the middle and 1 MCM bed on the left and right loop side, respectively. The magnetic reluctance $\mc{R}$ of the components are denoted by $\mc{R}_M$, $\mc{R}_L$ and $\mc{R}_R$, accordingly. The colors blue and red, correspond to a cold and hot MCM bed, respectively. Magnetic flux $\Phi$ is flowing in the direction indicated by the black arrows and distributed in the two loops depending on the thermomagnetic state of the two MCM beds. Two coils of $N$ turns are wound on each loop and connected in series such that the induced current $I$ is equal in both. Note that the product $NI$ corresponds to a magnetomotive force $\mc{F}$. Top: The electric circuit is composed of the two coils each with impedance $Z_{\text{coil}}$ and are connected to a load with impedance  $Z_{\text{load}}$. The induced current is flowing in the direction of the yellow arrows and is driven by the electromotive force $\mc{E}$ induced by the flux changes in the magnetic circuit. The two circuits are thus coupled by the two coils, illustrated with the dashed grey lines. Image of packed spheres were taken from Ref.\cite{Fathiganjehlou2024Multi-scaleBed}.}\label{magnetic_electric_circuit}
\end{figure}

\noindent We consider a typical thermomagnetic energy harvester with a figure-of-eight shaped geometry, as the experimental setup from Ref. \cite{Bahl2024DesignHarvester}, shown in Fig. \ref{Experimental_setup}. A diagram of the magnetic and electric circuits of such a typical TMG is shown in Fig. \ref{magnetic_electric_circuit}. 

\subsection{The magnetic circuit}
We begin by analyzing the magnetic circuit. The system can be considered as a closed magnetic circuit with three branches forming two loops. The topology consists of MCMs packed into a permeable bed on each side loop, a PM with a remanent flux density $B_{\text{\scriptsize rem}}$ located in the center branch, and soft iron yokes that connect the components and guide the magnetic flux through the circuit. Completing the circuit, an induction coil with $N$ winding turns is wound around each loop to harvest the induced emf generated by variations in magnetic flux. We assume that the coils are connected in series and wound in the same direction. 
\noindent Analogous to an electric circuit, the magnetic flux $\Phi$ circulating in the loops, indicated by the black arrows in Fig. \ref{magnetic_electric_circuit}, plays the role of an electric current, while the magnetomotive force (mmf), $\mc{F}$, serves as the magnetic equivalent of the electromotive force (emf). Following the magnetic circuit terminology, the magnetic flux is given by $\Phi = B A$, where $B$ is the magnetic flux density and $A$ is the cross-sectional area of the magnetic component. The mmf is defined as $\mc{F} = H l$, where $H$ is the magnetic field and $l$ is the length of the component. In linear magnetic materials, the magnetic flux density is proportional to the magnetic field and scales with the magnetic permeability $\mu$, according to $B = \mu H$. For permanent magnets, the remanent flux density $B_{\text{rem}}$ adds to this relation, giving $B = B_{\text{rem}} + \mu H$. With this terminology, and by analogy to an electric circuit, the behavior of the magnetic circuit can be described by Hopkinson’s law, the magnetic counterpart to Ohm’s law,
\begin{equation}
    \mc{F} = \mc{R} \Phi  
    \label{magnetomotive_force}
\end{equation}
where $\mc{R}$ is the magnetic reluctance of the magnetic components, which is given by
\begin{equation}
    \mc{R} = \frac{l}{\mu_0\mu_r A}
    \label{eq:reluctance}
\end{equation}
where $\mu_0$ is the magnetic permeability of the vacuum, and $\mu_r$ is the relative permeability of the corresponding magnetic component. For clarity, the $\mc{F}, \mc{R}$, and $\Phi$  symbols corresponding to the different magnetic components in Fig. \ref{magnetic_electric_circuit} are identified by the subscript letters ``L'' and ``R'' for the left MCM bed and right MCM bed, respectively. The permanent magnet is labeled with the subscript ``M''. Following the circuit analogy, Kirchhoff’s loop rule can be applied to derive the governing equations for the left and right loops of the magnetic circuit as
\begin{eqnarray}
    \mc{R}_M (\Phi_M-\Phi_{\text{\scriptsize rem}}) + \mc{R}_L \Phi_L &=& +I N \nonumber\\ \mc{R}_M (\Phi_M-\Phi_{\text{\scriptsize rem}}) + \mc{R}_R \Phi_R &=& -I N 
    \label{eq:K2}
\end{eqnarray}

where $I$ is the electric current, equal in magnitude through both coils since they are connected in series. The sign difference on the right-hand sides are due to the magnetic fluxes $\Phi_L$ and $\Phi_R$ circulating in opposite directions, while the coils are wound in the same direction so that the flux changes $d\Phi_L/dt$ and $d\Phi_R/dt$ will add constructively. Assuming that stray fields can be neglected, the total magnetic flux is conserved in the circuit, such that the sum of the left and right loop fluxes equals the flux provided by the PM, $\Phi_L + \Phi_R = \Phi_M$. Using this relation together with Eq. \eqref{eq:K2}, the magnetic circuit can be solved and the fluxes determined as
\begin{eqnarray}\label{eq:Flux}
    \Phi_L &=& \frac{\Phi_{\text{\scriptsize rem}} \mc{R}_M \mc{R}_R- I N (2 \mc{R}_M + \mc{R}_R)}{\mc{R}_L \mc{R}_M + \mc{R}_L \mc{R}_R + \mc{R}_M \mc{R}_R}, \nonumber\\  \Phi_R &=& \frac{\Phi_{\text{\scriptsize rem}} \mc{R}_L \mc{R}_M + I N (  2 \mc{R}_M + \mc{R}_L)}{\mc{R}_L \mc{R}_M + \mc{R}_L \mc{R}_R + \mc{R}_M \mc{R}_R} 
\end{eqnarray}

We note that the magnetic circuit solution depends on both the current $I$ and the number of coil turns $N$, which are properties of the induction coils. 

\subsection{The electric circuit}
\noindent We now turn to analyzing the electric circuit. The electric circuit, shown on the top of Fig. \ref{magnetic_electric_circuit}, consists of two coils connected in series with a load. We define $R_{\text{\scriptsize coil}}$  as the resistance of each coil and $R_{\text{\scriptsize load}}$ as the resistance of the external load. In addition, we introduce the possibility of adding a load capacitance, $C_{\text{load}}$ to the circuit, which has not yet been considered in previously numerical models that couples the magnetic and electric circuits of a TMG \cite{Jiang2022NumericalRecovery, Brillouin1948ThermomagneticGenerator}. The reason for adding a load capacitance to the circuit will be clarified when we later derive the expression for the maximum power output. With the electrical components of the circuit defined, Kirchhoff’s voltage law gives
\begin{equation}
    \left( 2R_{\text{coil}}+R_{\text{load}}\right)I+\frac{Q}{C_{\text{load}}} = - N (\frac{d\Phi_L}{dt} - \frac{d\Phi_R}{dt})    
    \label{Electrical_cicuit}
\end{equation}
where $Q = \int Idt$ is the electrical charge and the term $Q/C_{\text{load}}$ is the voltage drop across the load capacitor. The total resistance is given by $R_{\text{\scriptsize tot}} = 2R_{\text{coil}}+R_{\text{load}}$. The term on the right-hand side in Eq. \eqref{Electrical_cicuit} corresponds to the emf $\mc{E}$ induced in the two coils, as described by Faraday’s law of induction, Eq. \eqref{faradays_law}. In principle, it would be possible to have two independent electrical circuits, with the corresponding coil wound around the left and right loops of the magnetic circuit, respectively, and each connected to its own load. When describing such a configuration, Eq. \eqref{Electrical_cicuit} would be split into 2 electrical circuit equations, where each coil is coupled to the same magnetic circuit. By inserting the solution from the magnetic circuit, Eq. \eqref{eq:Flux}, into the electric circuit equation, Eq. \eqref{Electrical_cicuit}, we obtain a time-dependent differential equation that governs the current $I(t)$ induced in the coils. Solutions of $I(t)$ can be used to compute the harvested power of the system. However, we note that this  differential equation is only numerically solvable, and further simplifications are needed to solve the equation analytically.  

\subsection{Time dynamics and temperature change}
In a TMG the MCM experiences a time-varying temperature, through interaction with the oscillating hot/cold fluid stream. This means that one must account for the temperature-dependent magnetic properties of the MCMs. However, first it is of interest to discuss from where the available energy for harvesting originates. Starting from the differential form of Faraday’s law, $\bs{\nabla}\times \bs{E} = -\frac{d \bs{B}}{d t}$, where $\bs{E}$ denotes the electric field, we first multiply both sides of Faraday's law by the magnetic field $\bs{H}$
\begin{equation}
    \bs{H}\cdot \bs{\nabla}\times \bs{E} = - \bs{H}\cdot \frac{d \bs{B}}{d t}
\end{equation}
We then apply a standard vector identity to expand the left-hand side
\begin{equation}
\bs{\nabla}\cdot (\bs{E}\times\bs{H})    -\bs{E}\cdot \bs{\nabla}\times \bs{H} = - \bs{H}\cdot \frac{d \bs{B}}{d t}
\end{equation}
Realizing that Ampere's law, $\bs{\nabla}\times \bs{H} = \bs{J}$, where $\bs{J}$ is the current density, can be inserted in the second left hand side term, we obtain
\begin{equation}
\bs{\nabla}\cdot (\bs{E}\times\bs{H})    -\bs{E}\cdot \bs{J} = - \bs{H}\cdot \frac{d \bs{B}}{d t}
\label{without_omega}
\end{equation}
To derive a relation for the electromagnetic energy transfer in the TMG system, we note that the first term corresponds to the divergence of the Poynting vector, $\bs{S} = \bs{E}\times\bs{H}$, which represents the rate of energy flow. We now define a fixed arbitrary region $\Omega$ that fully encloses all the magnetic and electric components of the TMG, and we assume that the region is large enough such that all the electromagnetic fields vanish on and outside its boundaries. Integrating all terms in Eq. \eqref{without_omega} over this region then gives
\begin{equation}
-\int_{\Omega}  \bs{\nabla}\cdot (\bs{E}\times\bs{H}) dV    +\int_{\Omega} 
 \bs{E}\cdot \bs{J} dV = \int_{\Omega}  \bs{H}\cdot \frac{d \bs{B}}{d t} dV
 \label{Poynting}
\end{equation}
Since no net electromagnetic energy enters or leaves the boundary of $\Omega$, we apply the divergence theorem to the first term on the left hand side, which thus becomes the flux of the Poynting vector through the boundary of $\Omega$, where $\bs{S}$ is assumed to be zero. This simplifies Eq. \eqref{Poynting} to 
\begin{equation}
 \int_{\Omega} 
 \bs{E}\cdot \bs{J} dV = \int_{\Omega}  \bs{H}\cdot \frac{d \bs{B}}{d t} dV
 \label{eq:available_energy}
\end{equation}
Thus, the rate of energy dissipation in the coils through Joule heating, $\bs{E}\cdot \bs{J}$, is directly related to the rate of change of magnetic energy in the system, represented by $\bs{H}\cdot \frac{d \bs{B}}{d t}$, which both reflects the magnetic hysteresis behavior of the materials and the magnetic work of the MCMs. We note that the PM does not undergo a hysteresis cycle and therefore does not directly contribute to the net energy exchanged over one full cycle. However, its presence significantly influences the $\bs{H}$-field, thereby shaping the hysteresis loop experienced by the MCMs. As a result, it indirectly affects the total magnetic energy available for harvesting. Thus, the time dependent power being dissipated in the resistive parts of the electrical circuit corresponds exactly to the instantaneous rate of variation of the magnetic energy associated with the MCM beds. \\

Regarding the temperature-dependent magnetic properties of the MCMs, we here approximate the effect of thermal cycling by imposing a periodic, sinusoidal variation in the permeability of the MCMs, instead of modeling the full heat transfer process between the fluid and the MCM beds, as this among others requires a detailed knowledge of the internal geometry of the MCMs.  We consider a cycle in which the permeabilities of the left and right MCM beds vary sinusoidally in antiphase around a mean value $\overline{\mu}$, with an amplitude $\Delta_{\mu}$, which will depend on the MCM material used. In the following, the MCM beds are assumed to behave as linear magnetic materials, such that $\bs{B} = \mu \bs{H}$, where the permeability $\mu$ varies only with temperature and is independent of the magnetic field strength. Later the magnetic dependency is introduced as magnetic saturation. Under these assumption, the reluctances $\mc{R}_L$ and $\mc{R}_R$ become time-dependent and are given by
\begin{eqnarray}
     \mc{R}_L (t) = \frac{l_{\text{\scriptsize bed}}}{A_{\text{\scriptsize bed}}(\overline{\mu}+\Delta_{\mu}\sin(\omega t))}, \nonumber\\ \mc{R}_R (t) = \frac{l_{\text{\scriptsize bed}}}{A_{\text{\scriptsize bed}}(\overline{\mu}-\Delta_{\mu}\sin(\omega t))}
     \label{R(t)}
\end{eqnarray}

\noindent where $\omega$ is the angular frequency of the oscillation, and $l_{\text{\scriptsize bed}}$ and $A_{\text{\scriptsize bed}}$ are the length and cross-sectional area of the MCM beds, respectively. With the imposed sinusoidal variation of the permeability, $\overline{\mu}-\Delta_{\mu}$ and $\overline{\mu}+\Delta_{\mu}$ correspond to a warm and cold MCM bed, respectively. Inserting Eq. \eqref{R(t)} into the solutions to the magnetic circuit, Eq. \eqref{eq:Flux}, the flux difference can be determined as
\begin{equation}\label{eq:FluxDiff01}
    \Phi_L-\Phi_R = \ell(t) I -\Delta_{\Phi}\sin(\omega t)
\end{equation}
where $\ell$ is a time-dependent parameter and $\Delta_{\Phi}$ is a time-independent parameter, respectively defined as 

\begin{eqnarray}
   \ell(t) &=& -N\left(\frac{2\mc{R}_M}{\overline{\mc{R}}_{\text{\scriptsize bed}}}\right)\cdot{}\nonumber\\&&\left(\frac{1 + \overline{\mc{R}}_{\text{\scriptsize bed}}/(2\mc{R}_M ) - (\Delta_{\mu}/\overline{\mu})^2\sin^2(\omega t)}{\mc{R}_M + \overline{\mc{R}}_{\text{\scriptsize bed}}/2}\right)
   \nonumber\\ \Delta_{\Phi} &=& - \Phi_{\text{\scriptsize rem}} \left(\frac{\Delta_{\mu}}{\overline{\mu}}\right)\left( \frac{\overline{\mc{R}}_{\text{\scriptsize bed}}}{2 \mc{R}_M} + 1 \right)^{-1}
    \label{eq:ell_Dphi}
\end{eqnarray}

\noindent Here we introduced the average reluctance of the MCM beds, $\overline{\mc{R}}_{\text{\scriptsize bed}}$ defined as
\begin{equation}
\overline{\mc{R}}_{\text{\scriptsize bed}} = \frac{l_{\text{\scriptsize bed}}}{A_{\text{\scriptsize bed}}\overline{\mu}}
\end{equation}
which corresponds to the reluctance of the MCM bed when $\mu = \overline{\mu}$.
\noindent Differentiating Eq. \eqref{eq:FluxDiff01} with respect to $t$, multiplying by $N$ and noticing that $\Delta_{\Phi}$ is time-independent, the total emf $\mc{E}$ is given by
\begin{eqnarray}
    \mc{E} &=& N (\frac{d\Phi_L}{dt} - \frac{d\Phi_R}{dt}) \nonumber\\ &=& N\frac{d\ell(t)}{dt}I(t) + N \ell(t) \frac{dI(t)}{dt} - N\omega \Delta_{\Phi} \cos(\omega t)
    \label{emf}
\end{eqnarray}
Upon substitution of the above equation in Eq. \eqref{Electrical_cicuit}, we obtain a linear ordinary differential equation (ODE) with time-dependent coefficients having the current, $I(t)$ as unknown. Inserting this expression of $\mc{E}$, Eq. \eqref{emf}, into the electrical circuit, Eq. \eqref{Electrical_cicuit}, we obtain the equation for the coupled system 
\begin{eqnarray}
    \left(R_{\text{tot}}+N\frac{d\ell(t)}{dt}\right) I(t) + N\ell(t)\frac{dI(t)}{dt}+\frac{Q(t)}{C_{\text{load}}}-N\omega \Delta_{\Phi}\cos(\omega t) && \nonumber \\ = 0 &&
    \label{eq:Coupled_system_ODE}
\end{eqnarray}
In its present form, this equation must be solved numerically, with the current expressed as the time derivative of the charge $Q$, i.e. $I(t) = \frac{dQ(t)}{dt}$ and $\frac{dI(t)}{dt} = \frac{d^2I(t)}{dt^2}$.   

\subsection{Analytical approximation}
An analytical solution will be more transparent and assessing the solution in closed form would provide a better understanding of the physics and impact that the different parameters, $N$, $\omega$ etc., have on the system performance. 
However, in order to solve Eq. \eqref{eq:Coupled_system_ODE} analytically, we need to introduce one approximation. Inspecting the definition of $\ell(t)$ in Eq. \eqref{eq:ell_Dphi}, we notice that, as long as the amplitude of the oscillating term is small, we can treat $\ell(t)$ as time-independent and constant. This condition is fulfilled when 

\begin{equation}
    \left(\frac{\Delta_{\mu}}{\overline{\mu}}\right)^2   \ll \left(1+\frac{1}{2}\frac{\overline{\mc{R}}_{\text{\scriptsize bed}}}{\mc{R}_M}\right) \label{condition2}
\end{equation} 

where it should be noted that the left side term will always be between zero and one. When the above condition is met, we are justified in dropping the oscillating term from the definition of $\ell$. This corresponds to evaluating the time-dependent parameters at time $t = 0$, which corresponds to the moment when the permeability of the left and right MCM bed are both equal to each other and to the average value $\overline{\mu}$. With this approximation of $t = 0$, the two time-dependent parameters $\frac{d\ell(t)}{dt}$ and $\ell$ become
\begin{equation}
   \frac{d\ell(t=0)}{dt} = 0,   \quad \ell(t = 0) = 
    -\frac{2 N}{\overline{\mc{R}}_{\text{\scriptsize bed}}}
    \label{l(0)}
\end{equation}
Inserting Eq. \eqref{l(0)} into Eq. \eqref{emf}, it is seen that the term $N\ell(0)$ corresponds to a self-inductance $L$ defined as
\begin{equation}\label{eq:selfLapprox}
    L = -N \ell(t = 0) = 2\frac{N^2}{\overline{\mc{R}}_{\text{\scriptsize bed}}}
\end{equation}
Therefore, in the coupled circuit, the self-inductance $L$ in the electric circuit will align with the magnetic circuit if $L$ is defined as Eq. \eqref{eq:selfLapprox}. Note that this definition of $L$, which is the proportionality factor between the emf and $\frac{dI}{dt}$, is usually inversely proportional to the total reluctance of the magnetic circuit, which the coil is associated with\cite{Umans2014FitzgeraldMachinery}. However, our analysis reveals that the best approximation of a time-independent self-inductance only associates with the average reluctance of the MCM bed. With the introduced approximation the coupled circuit equation can be written as,
\begin{equation}
        R_{\text{\scriptsize tot}}I(t)- L\frac{dI(t)}{dt} +\frac{Q(t)}{C_{\text{load}}} -N\omega \Delta_{\Phi}\cos(\omega t) = 0 
        \label{Final ODE}
\end{equation}
We now introduce the phasor notation to simplify the equation further, where the current $\tilde{I}$, total impedance $\tilde{Z}_{\text{tot}}$ and the emf $\tilde{\mc{E}}$ are defined respectively as,
\begin{eqnarray}
    \tilde{I} &=& |I|e^{i(wt-\phi)} \nonumber \\  \tilde{Z}_{\text{tot}} &=& \left(R_{\text{\scriptsize tot}}+\frac{1}{i\omega C_{\text{load}}}-i\omega L\right) e^{i\phi}  \nonumber \\  \tilde{\mc{E}} &=& N\omega\Delta_{\Phi}e^{i\omega t}
\end{eqnarray}
where $\phi = X_{\text{\scriptsize tot}}/R_{\text{\scriptsize tot}}$ is the impedance phase. With this notation, the steady-state current amplitude, $|I|$, can now be analytically calculated as 
 
\begin{equation}
    |I| = \frac{|\mc{E}|}{|Z_{\text{\scriptsize tot}}|} \label{current_amplitude} = \frac{\left|N\omega\Delta_{\Phi}\right|}{\left|R_{\text{\scriptsize tot}}+\frac{1}{i\omega C_{\text{load}}}-i\omega L\right|}
\end{equation}

Returning to the time-domain representation, the current, $I(t)$, can be defined by taking the real part of $\tilde{I}$, 
\begin{equation}
    I(t) = \Re(\tilde{I}) = |I| \cos(\omega t - \phi)
    \label{time-dependent_current}
\end{equation}

\noindent The average power $\mc{P}_{\text{avg}}$ dissipated in the load is then given by,
\begin{equation}
    \mc{P}_{avg} = \frac{1}{\tau}\int_0^{\tau} R_{\text{load}}I(t)^2 dt
    \label{eq:Avg_Power}
\end{equation}
where $\tau = 2\pi / \omega$ is the period of one cycle. 
The maximum power transfer theorem\cite{Zeng2021MaximumTransfer} states that the power is maximized when the load and coil impedances are complex conjugates, such that $Z_{\text{\scriptsize coil}} = Z_{\text{\scriptsize load}}^*$. This is achieved when $C_{\text{load}} = 1/L\omega^2$ and $R_{\text{load}} = 2R_{\text{coil}}$, which yields a phase of $\phi = 0$. Combining Eqs. \eqref{time-dependent_current} and \eqref{eq:Avg_Power}, the average power with matched impedance is given as
\begin{equation}
    \mc{P}_{avg} = \frac{N^2 \omega^2 \Delta_{\Phi}^2}{16 R_{\text{\scriptsize coil}}}
    \label{eq:power_matched}
\end{equation}
The above expression is different from Eq. (13) derived in Jiang et al. \cite{Jiang2022NumericalRecovery}, due to the fact that we in this work have included a load capacitor in the electric circuit. In the case where no load capacitor is present, we would have obtained another expression for the optimal load resistance, $R_{\text{load}} = \sqrt{4R_{\text{coil}}^2+L^2\omega^2}$, which is similar to the one obtained by Jiang et al. Following the same analytical derivation as above, but this time without a load capacitor, we obtain the expression for the average power 
\begin{equation}
   \mc{P}_{\text{avg}} = -\frac{\Delta_{\Phi}^2 N^2 \left( 2R_{\text{coil}}-\sqrt{L^2w^2+4R_{\text{coil}}^2}\right)}{4L^2}
   \label{eq:Power_withoutLoadCapacitor}
\end{equation}

\subsubsection{Including the coil volume}
If the available copper volume for winding a coil is fixed, an important question arises, namely how many turns $N$ should be made, and what should the wire radius $r_w$ be to maximize the power output of a TMG? The DC resistance of a long coil is given by
\begin{equation}
    R_{\text{\scriptsize coil}} = \rho \frac{ 2\pi a N}{\pi r_w^2} = \rho \frac{2 a N}{r_w^2} 
    \label{eq:Rcoil}
\end{equation}
where $a$ is the winding radius and $\rho$ is the resistivity of the wire. The above expression corresponds to the DC resistance of the coil and neglects frequency dependent effects such as skin effects. This approximation is justified by the fact that TMGs typically operate at frequencies on the order of 1 Hz or lower. The number of turns can be written as a function of the coil volume $V_{\text{coil}} = 2\pi a N \pi r_w^2$ as 
\begin{equation}
    N = \frac{V_{\text{coil}}}{2\pi^2 a r_w^2}
    \label{eq:N}
\end{equation}
Substituting this expression of $N$ and Eq. \eqref{eq:Rcoil} into Eq. \eqref{eq:power_matched} we obtain a volume dependent relation of the average power, 
\begin{equation}
    \mc{P}_{\text{avg}} = \frac{V_{\text{coil}} \Delta_{\phi}^2 \omega^2}{64 \pi^2 a^2 \rho}
    \label{eq:power_V}
\end{equation}
where $\mc{P}_{\text{avg}}$ is now expressed in terms of $V_{\text{coil}}$ instead of $N$ and $r_w$, revealing that the power does not depend on the specific combination of $N$ and $r_w$, but only on the volume $V_{\text{coil}}$. We note that in this simplified analytical model, the power scales indefinitely with $V_{\text{coil}}$, which is not always realistic, as discussed subsequently. 

Repeating the above analytical derivation, but instead inserting Eqs. \eqref{eq:Rcoil}, \eqref{eq:N} and $R_{\text{load}} = \sqrt{4R_{\text{coil}}^2+L^2\omega^2}$ into Eq. \eqref{eq:Power_withoutLoadCapacitor}, which is the expression for the average power without a load capacitor, we obtain the average power
\begin{equation}
   \mc{P}_{\text{avg}} =  -\frac{\Delta_{\Phi}^2V_{\text{coil}}^2\left(2V_{\text{coil}}\rho-\sqrt{L^2\pi^4r_w^8\omega^2+4V_{\text{coil}}^2\rho^2}\right)}{16L^2a^2\pi^6r_w^8}
   \label{eq:Power_withoutLoadCapacitor_V}
\end{equation}
The above expression of the average power without a load capacitor has a non-linear dependency on the coil volume $V_{\text{coil}}$ and wire radius $r_w$. However, Eq. \eqref{eq:Power_withoutLoadCapacitor_V} can still be used to compute the average power of an experimental TMG without a load capacitor, which will be shown further below.

\subsubsection{Optimizing the magnetic components}
\noindent While the maximum power dissipated in the load resistance of the electric circuit can be optimized using the maximum power transfer theorem, the analogous result applicable to magnetic circuits is not as well-known. Effective magnetic circuit design aims to maximize flux linkage to the coils, minimize flux leakage, and ensure a high reluctance contrast between the hot and cold MCM bed to enable substantial flux variations during the thermal cycling. As shown in the previous section, the average harvested power, $\mc{P}_{\text{avg}}$, is proportional to the square of $\Delta_{\Phi}$, which represents the time-independent parameter that scales the amplitude of the oscillation defined in Eq. \eqref{eq:ell_Dphi}. Since $\Delta_{\Phi}$ depends on the geometric and magnetic properties of both the PM and the MCM beds, this suggests that an optimal ratio between their reluctances may exist.

To explore this further, we express the reluctance $\mc{R}$ in terms of geometry using $V_M = A_M l_M$ and rearranging Eq. \eqref{eq:reluctance}. We thus get  
\begin{equation}
    A_M = \left( \frac{V_M}{\mu_M \mc{R}_M}  \right)^{1/2}, \,\ l_M = \left(V_M\mu_M \mc{R}_M \right)^{1/2}, \,\ \Phi_{\text{rem}} = A_M B_{\text{rem}}
    \label{relations2}
\end{equation}
with the corresponding expressions for the MCM beds being identical to the two first expressions. The last term is simply the relation between remanent flux and remanent flux density. Substituting these expressions into Eq. \eqref{eq:power_V}, and setting the derivative with respect to $\mc{R}_M$ equal to zero, we obtain

\begin{equation}
    \frac{d \mc{P}_{\text{avg}}}{d \mc{R}_M} = 0 \Rightarrow \mc{R}_M = \frac{\overline{\mc{R}}_{\text{bed}}}{2}
    \label{optimal_reluctance}
\end{equation}
The derived relation indicates that maximum power output is achieved when the reluctance of the PM, $\mc{R}_M$, equals half of the average reluctance of the MCM beds, $\overline{\mc{R}}_{\text{bed}}$, which is also equal to the combined average reluctance of the two beds that are connected in parallel. This condition resembles the impedance matching principle in the electric circuit, where maximum power transfer occurs when the load resistance equals the internal resistance of the source. Inserting the optimal PM reluctance, Eq. \eqref{optimal_reluctance} into the power equation, Eq. \eqref{eq:power_V}, we obtain

\begin{equation}
    \mc{P}_{\text{avg}} =  \frac{B_{\text{rem}}^2\omega^2 V_{\text{coil}} V_M  }{32\pi^2 a^2 \rho \overline{\mc{R}}_{\text{bed}} \mu_M} \left(\frac{\Delta_{\mu}}{\overline{\mu}}\right)^2
    \label{Power_optimized_reluctance}
\end{equation}
The above equation represents the analytical expression for the optimized power output. We observe that the power scales linearly with the magnet volume $V_M$ and quadratically with the remanent magnetic flux density $B_{\text{rem}}$. We note that the power does not explicitly depend on the volume of the MCM beds, but rather on their average reluctance, $\overline{\mc{R}}_{\text{bed}}$.  However, it is important to emphasize that Eq. \eqref{Power_optimized_reluctance} is only accurate as long as the conditions of Eqs. \eqref{condition2} and \eqref{optimal_reluctance} are satisfied, whereas Eq. \eqref{eq:Coupled_system_ODE} always applies although this has to be solved numerically.

\subsection{Magnetic materials with saturation}
Returning to the non-simplified model for the TMG, Eq. \eqref{eq:Coupled_system_ODE}, it is useful to consider magnetic materials that display a saturation in permeability with field, as this is a more realistic model of actual MCM materials. We incorporate the non-linear magnetic properties by accounting for the material’s saturation magnetization $M_s$ within the $B$-$H$ constitutive relation. This non-linear behavior is governed by two key parameters: the time-dependent permeability $\mu$, which determines the slope of the $B$-$H$ curve before saturation, and $M_s$, beyond which the material response becomes linear with slope $\mu_0$. Although the precise form of the $B$-$H$ curve can be obtained through experimental characterization of the specific MCM, we approximate it using a piecewise linear model. In this model, we assume that for $|M| > M_s$, the slope transitions to $\mu_0$. This behavior can be mathematically described as,
\begin{equation}\left\{
    \begin{array}{l l }
         B = \mu_0 H - B_0,&\text{ for } H<-H_0\\
         B = \mu H,&\text{ for }-H_0<H<+H_0  \\
         B = \mu_0 H + B_0,&\text{ for } H>+H_0
    \end{array}\right.
    \label{BH-relation}
\end{equation}
where the two parameters $H_0$ and $B_0$ are the values of field and flux density corresponding to the onset of saturation, and are computed from $\mu$, $\mu_0$ and $M_s$. This piece-wise linear material model is only included in the numerical solution of Eq. \eqref{eq:Coupled_system_ODE}.

\section{Validation of the model}
To validate the presented model, we consider a TMG identical to the experimental prototype described in Ref. \cite{Bahl2024DesignHarvester}, i.e. with the following properties of $N = 156$, $\omega = \pi$ Hz, $r_w = 0.9$ mm, $a = 50$ mm, hot reservoir temperature $T_{\text{H}} = 303$ K and cold reservoir temperature $T_{\text{C}} = 283$ K. This is the device shown in Fig. \ref{Experimental_setup}.

\begin{figure}[!ht]
    \centering
    \includegraphics[width=0.99\linewidth]{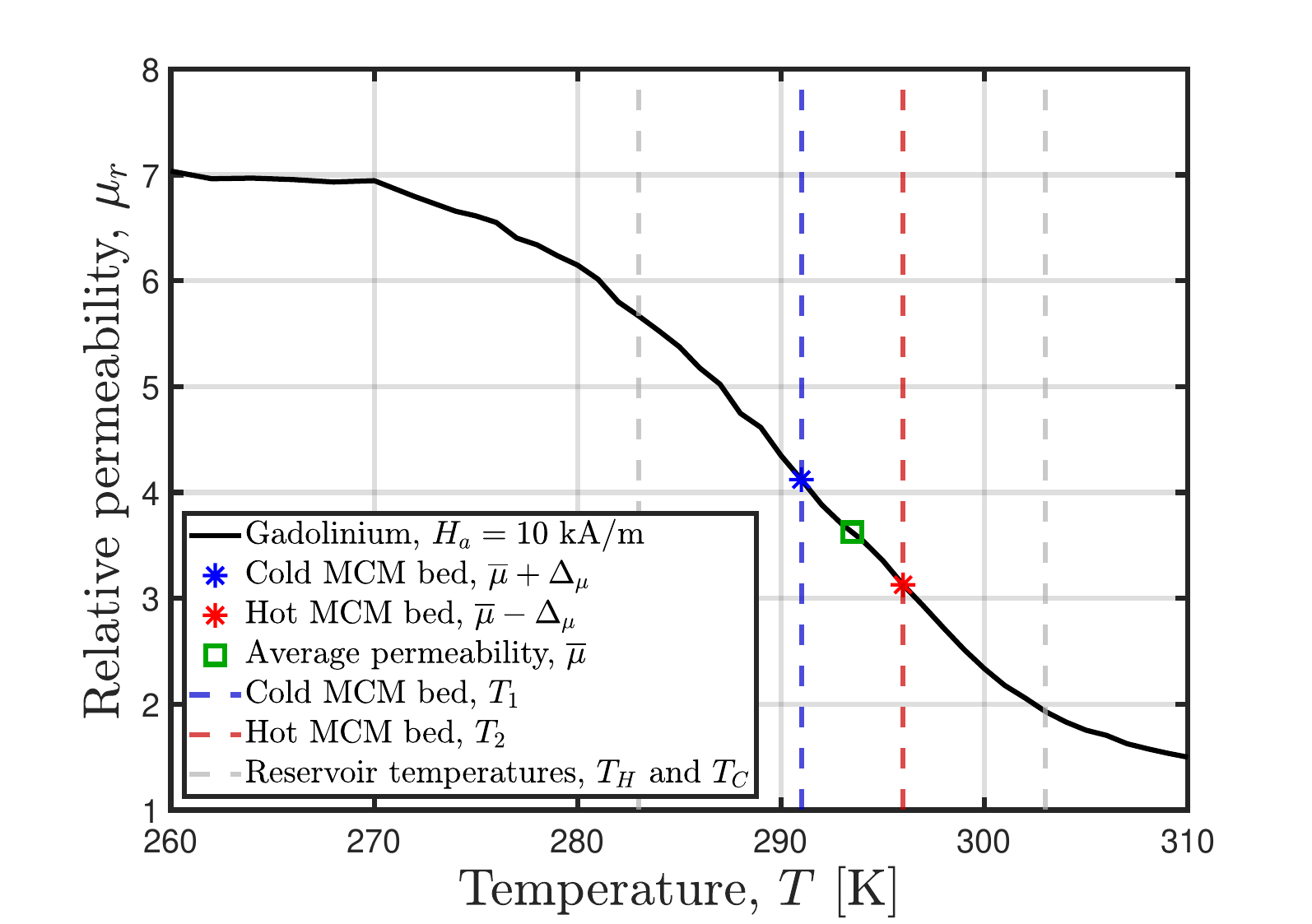}
    \caption{Experimental derived relative permeability of the gadolinium spheres, which are used in the TMG, as function of temperature in an applied field of $H_0 = 10$ kA/m. The blue and red dot indicate the relative permeability of a cold and hot MCM bed respectively. The cold and hot MCM temperatures are estimated with Eq. (5) from Ref. \cite{Almanza2017NumericalCycle}. $\overline{\mu}$ is indicated by a green square. The dashed grey lines shows the reservoir temperatures.}
    \label{mu_vs_temperature}
\end{figure}

To predict the performance of the TMG, we first determine the change in temperature of the MCM. 
To estimate the heat transferred to the MCM and the resulting temperature change, we employ the heat exchange model proposed by Almanza et al. \cite{Almanza2017NumericalCycle}
\begin{equation}
\delta Q = -\left[k_h(T(t)-T_h)+ k_c(T(t)-T_c)\right]dt
\label{heat2}
\end{equation}
where $k_h$ and $k_c$ are the thermal exchange constants of the hot and cold reservoirs, respectively, and $T_h$ and $T_c$ are the corresponding reservoir temperatures. As in Almanza et al. we assume constant exchange values of $k_h = k_c = 1$ W cm$^{-3}$ K$^{-1}$. This model is used with the properties of the TMG and with gadolinium magnetization data taken from Ref. \cite{Bjrk2010MagnetocaloricGd} as input. We estimate a temperature change of $\Delta T \sim 5$K, which is four times lower than the difference between the two reservoirs $\Delta T_{\text{res}} = 20$K. Fig. \ref{mu_vs_temperature} shows the relative permeability of gadolinium as function of the temperature based on the magnetization data input from Ref. \cite{Bjrk2010MagnetocaloricGd}. The green square shows the average permeability of $\overline{\mu} = 3.62$. The blue and red dot illustrate the relative permeability of gadolinium at the estimated temperatures of the cold and hot MCM beds respectively. Grey dashed lines illustrate the reservoir temperatures and thus the maximum permeability sweep available for the TMG.           

\begin{figure}[!ht]
    \centering
    \includegraphics[width=0.99\linewidth]{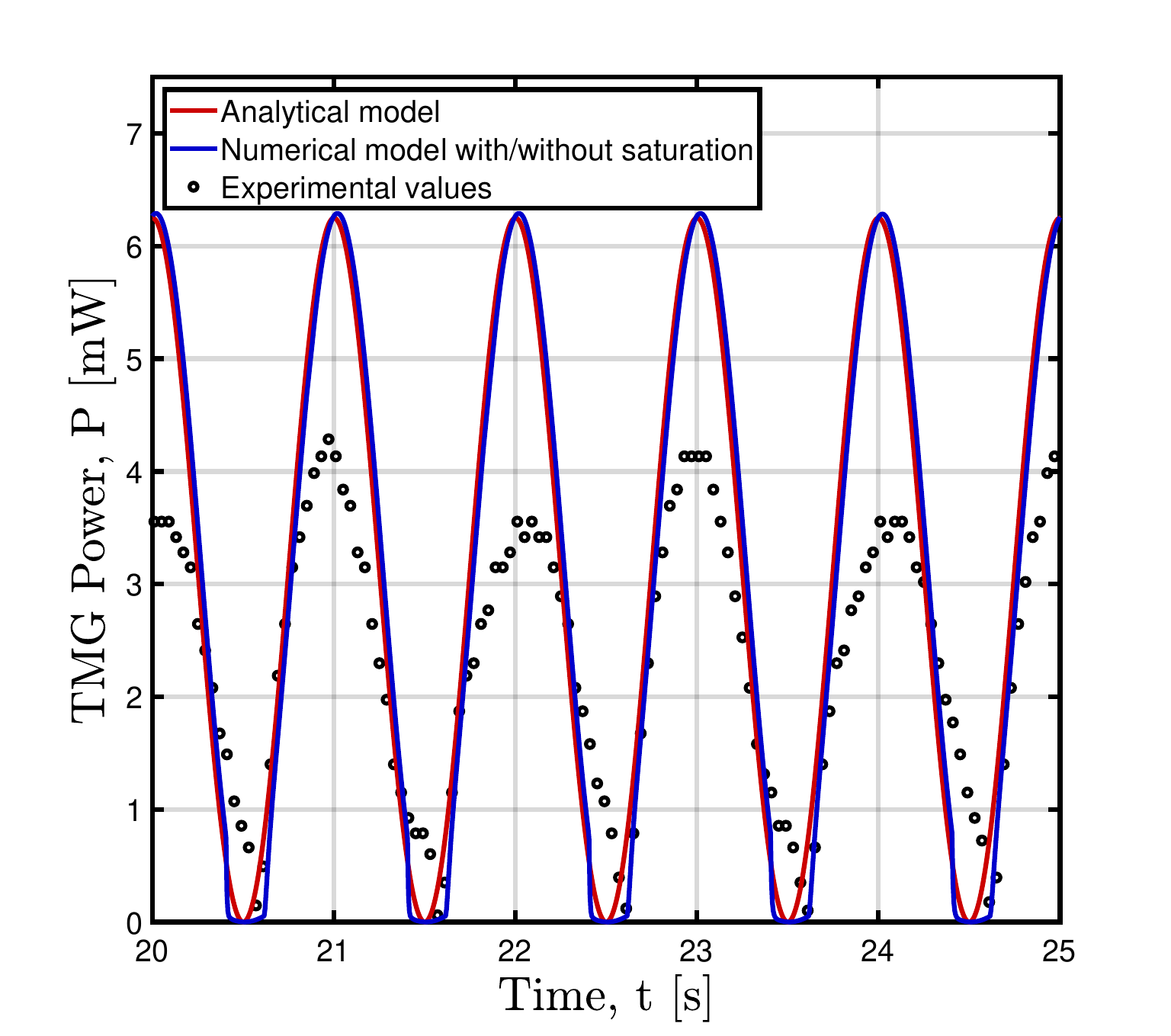}
    \caption{Power of the evaluated TMG from Ref. \cite{Bahl2024DesignHarvester} and with the numerical model, Eq. \eqref{eq:Coupled_system_ODE} and the analytical, Eq. \eqref{eq:power_V}}
    \label{InstantPvsTime_plot}
\end{figure}

The governing ODE, Eq. \eqref{eq:Coupled_system_ODE}, which must be solved numerically, has been implemented in Comsol Multiphysics for both a MCM material without and with a saturation magnetization described by Eq. \eqref{BH-relation}. We use both the analytical and the numerical model to calculate the time dependent power of the TMG as a function of time which is shown in Fig. \ref{InstantPvsTime_plot}. Similarly to the experimental results, the load resistance was chosen to fit the resistance of the two coils. We set the load capacitance to zero due to the fact that no load capacitor was present in the experimental setup. However, if this had been included and set to match the reactance of the coils then the power would have been enhanced according to the maximum power transfer theorem.     
Also shown in the figure are the experimental values from Ref. \cite{Bahl2024DesignHarvester} provided in Ref. \cite{Bahl2024DataHarvester}. The time interval was chosen such that the experimental setup had reached steady-state operation. The validation shows good agreement between experiment and the model with the exception that the experimental values are around 60\% lower in amplitude. This deviation is expected as the model presented here has no fitting parameters but relies only on experimentally measured properties and geometrical dimensions. The overestimation of the model can result from both thermal effects such as a too-high thermal exchange constants and flow maldistribution, from magnetic effects such as flux leakage and eddy currents, and from combined effects e.g. that the MCM beds does not undergo uniform and instantaneous temperature and magnetization changes. It is known that TMG models tend to overestimate the power output of the TMG devices 
 \cite{Jiang2022NumericalRecovery,mca26020047}. However, the trends gained from models are still accurate and models do provide an upper bound for the power that an ideal device could generate. 
 
 Additionally, we note that the analytical model and numerical model are seen to be very similar which clearly demonstrates the validity of the analytical approximation expressed by Eq. \eqref{condition2} for these parameters.      

\subsection{Optimizing power of a TMG}
Having demonstrated the agreement between the presented analytical and numerical models with the experimental data, we now consider how much power the system could have produced with optimal coils. 

\begin{figure}[!t]
    \centering
    \includegraphics[width=0.99\linewidth]{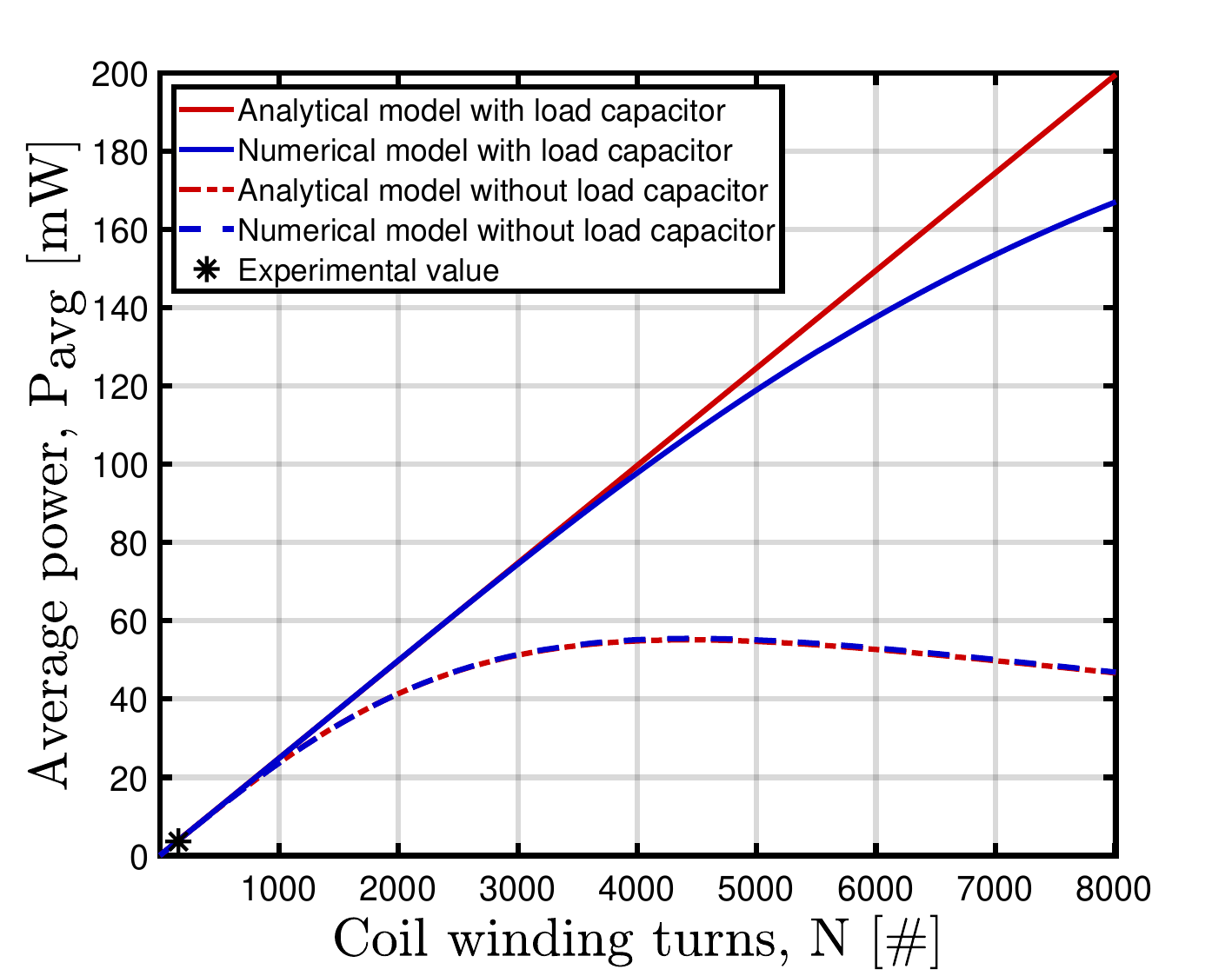}
    \caption{Average power of the evaluated TMG as function of coil turns $N$ for the experimental setup described in Ref. \cite{Bahl2024DesignHarvester}. The red and blue solid lines correspond to the analytical and numerical model, respectively, with a load capacitor. In addition, the red and blue dashed lines correspond to the analytical and numerical model without a load capacitor, respectively. Both numerical computations of the average power are with magnetic saturation of $M_s = 0.52$ T implemented. The average power of the experimental setup is indicated by the black star in the bottom left corner.}
    \label{Power_vs_N}
\end{figure}

Fig. \ref{Power_vs_N} shows the average power of the TMG as function of the number of coil winding turns. The red and blue solid lines correspond to the analytical and numerical model, respectively, with a load capacitor. In addition, the red and blue dashed lines correspond to the analytical and numerical model without a load capacitor, respectively. Both numerical computations of the average power are with magnetic saturation of $M_s = 0.52$ T implemented. For simplicity the numerical model without saturation was not shown, because they overlap the analytical model for all values of $N$, which confirms that the condition in Eq. \eqref{condition2} is satisfied for the chosen parameters. Observing Fig. \ref{Power_vs_N}, we first note that the analytical model with a capacitor is in fact unbounded in power as function of the number of coil turns as it should according to Eq. \eqref{eq:power_V} and that the numerical model with a capacitor deviates the analytical model for larger N values, which shows the effect of implementing magnetic saturation. Secondly, we note that the analytical and numerical model without a load capacitor match completely and reveal an optimum number of coil windings, $N$, where the average power is maximized. Since the experimental setup described in Ref. \cite{Bahl2024DesignHarvester} is without a load capacitor, we can directly observe in Fig. \ref{Power_vs_N} that changing the number of turns to the value of $N = 4500$ from the experimental value of $N=156$ would have enhanced the average power by a factor of 15. We stress, that the result of enhanced average power by a factor of 15, is without a load capacitor. Additionally, if a load capacitor was implemented the power could be enhanced and optimized even further to the value of 210 mW i.e. a factor of 58 by changing $N$ to the value of $N = 17000$. Furthermore, we observe in Fig. \ref{Power_vs_N} that without a load capacitor the gain in average power from adding more coil turns becomes smaller once $N > 1000$. This means that if increasing $N$ up to 4500 is too costly or impractical for the TMG design: considerable gains can still be obtained from a more moderate increase in $N$. This observation also describe the trend in average power with a load capacitor, but the gain becomes smaller once $N > 5000$ instead. As a final note, we stress that if the experimental TMG were to be optimized with $N>1000$, it is essential to first implement a load capacitor in the TMG design such that the trend of a gain in average power is both larger and sustained for larger values of $N$.

\begin{figure}[!t]
    \centering
    \includegraphics[width=1\linewidth]{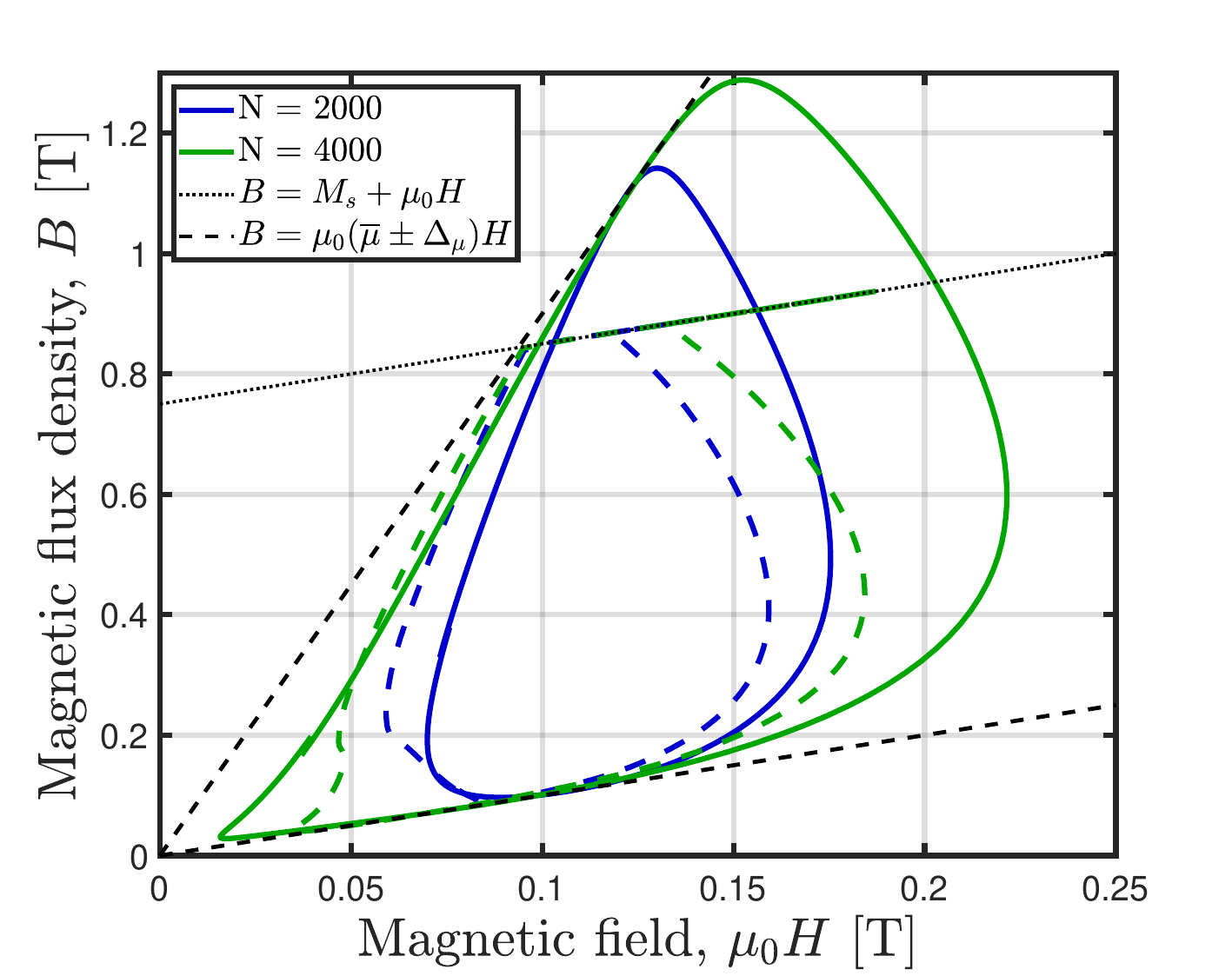}
    \caption{$B$-$H$ trajectories of the MCM beds over one operation cycle. blue and green correspond to $N = 2000$ and $N = 4000$ respectively. Solid and dashed curves denote the numerical model without and with saturation, respectively. All trajectories are bounded by the constitutive equation relations $\bs{B} = \mu_0\left(\overline{\mu} \pm \Delta_{\mu} \right)\bs{H}$, illustrated by the dashed black lines. In addition, the model with saturation is limited by, $\bs{B} = M_s + \mu_0 \bs{H} $, illustrated by the dotted black line.}
    \label{B_H_curve}
\end{figure}

Considering Eq. \eqref{eq:available_energy} from section 2.3, we remember that the available energy in the TMG is directly related to the hysteresis behavior of the MCM beds. Fig. \ref{B_H_curve} shows four MCM hysteresis loops. Solid and dashed curves correspond to the numerical model without and with saturation, respectively. The two colors blue and green correspond to $N = 2000$ and $N = 4000$ respectively. All curves are bounded by the $B$-$H$ constitutive equation of $B = \mu_0\left(\overline{\mu} \pm \Delta_{\mu} \right)H$, illustrated by the dashed black lines in the figure. In addition, the model with saturation is limited by, $B = M_s + \mu_0 H$, illustrated by the dotted black line in the figure. Since the available energy per cycle is determined by the area enclosed by the curve, the enclosed area of the dashed loops are limited by the saturation curve. This illustrates how implementing magnetic saturation in the model restricts the growth of the enclosed area, and thereby the available energy, as it is expected. 

\section{Evaluating existing TMGs}
The above finding, that the power produced by the TMG considered in Ref. \cite{Bahl2024DesignHarvester} could have been increased by a factor of 15-58 if larger coils had been used, might also apply to other prototype TMGs presented in literature. 

In the following, we consider TMG devices reported in literature and evaluate their potential increase in power output, had optimal coils been used. However, given the novelty of this scientific field, no standardized approach currently exist for designing thermomagnetic energy harvesters, making direct comparisons challenging. For each device, the reported geometric sizes of the magnetic component, winding radius and operating parameters were used as input to the model. When certain parameters were not specified in the literature, reasonable values were assumed as described below. The model was then used to calculate the average power as function of number of coil turns, following the procedure illustrated in Fig. \ref{Power_vs_N}. The optimal number of turns was identified as the value that maximized the average power, and the corresponding power output and coil turns was recorded as the optimized conditions with maximum achievable power for that device. We note that the evaluated TMGs have different operation frequency, reservoir temperatures and number of PMs, MCM beds and coils, meaning that their absolute power cannot directly be compared.

\begin{table*}[!t]
\centering
    \hskip-1.5cm \begin{tabular}{ |c|c|c|c|c|c|c|c|c| }
 \hline
 Study & MCM & $r_w$ [mm] & $N$ & $\omega$ [Hz] & $P_{exp}$ [mW] & $P_{max}$ [mW] & Optimal $N$  \\
 \hline
 Bahl et al. \cite{Bahl2024DesignHarvester} & Gd & 0.9 & 156 & $\pi$ & 3.6  & 210 & 17000  \\ 
 \hline
 Liu et al. \cite{Liu2023High-performanceSwitch} & Gd & 0.16* & 1600 & $\pi/2$ & 0.12  & 47 & 220000 \\
 \hline
 Dzekan et al. \cite{Dzekan2021CanGenerator} & Gd & 0.19*& 1000 & 2.2$\pi$ & 0.18 &  37  & 50500 \\
 \hline
 Waske et al. \cite{Waske2019EnergyTopology} &  LaFeCoSi & 0.19* & 1000 & 1.6$\pi$ & 1.2 & 14 & 23000 \\
 \hline

\end{tabular}
    \caption{The experimentally measured power and the potential maximum power, had optimal coils and an optimal load capacitor been used.  A "*" indicate values of wire radius, which was not given in the citation.  Dzekan\cite{Dzekan2021CanGenerator} and Waske\cite{Waske2019EnergyTopology} report the maximum experimental power output in their publications. }
    \label{tab:my_label}
\end{table*}

\begin{figure}[!b]
    \centering
    \includegraphics[width=0.9\linewidth]{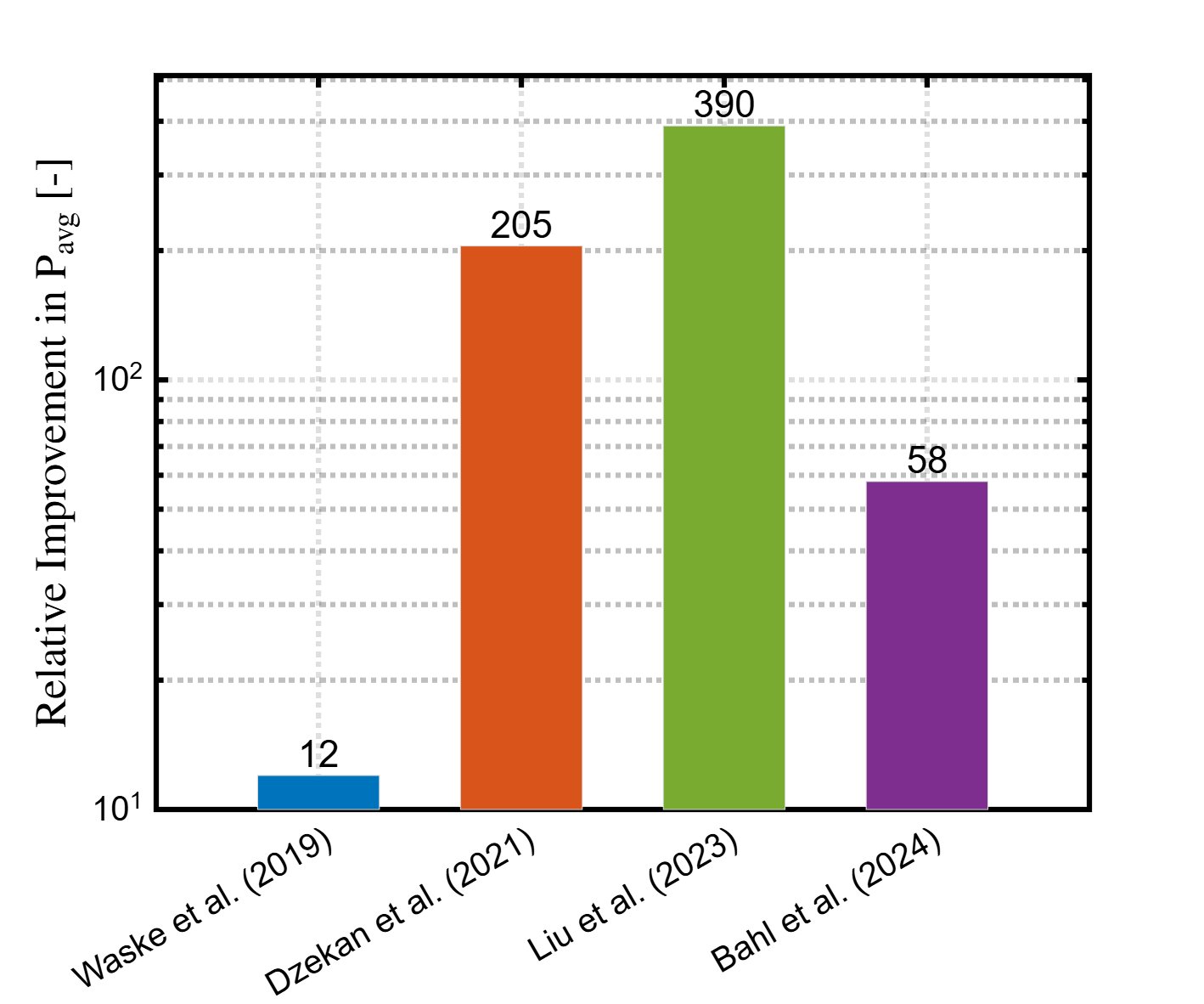}
    \caption{Relative improvement possible in average power predicted by the optimization of the four TMG devices evaluated. The bars indicate the ratio between the optimized power output predicted by the model and the experimentally reported power output. All devices show significant potential for improvements through coil optimization and implementation of an optimal load capacitance.}
    \label{bar_chart}
\end{figure}

\noindent Table \ref{tab:my_label} summarizes the experimental parameters and MCM materials of the evaluated TMG devices used in the model optimization. In addition, the table lists the average experimental power output reported in the literature and the maximum power output predicted by the model presented in this work, had optimal coils been used and with an optimal matched load capacitance.
It can be seen from the table that all evaluated TMG devices have the potential of increasing their average power output had the coils been larger. To illustrate this more clearly, Fig. \ref{bar_chart} presents a comparative bar chart showing the relative improvement in average power for each device.  For all devices the increase in power is at least a factor of 10, and for Liu et al. \cite{Liu2023High-performanceSwitch} and Dzekan et al. \cite{Dzekan2021CanGenerator} the increase in power is as large as a factor of 390 and 205, respectively. This large increase in power mainly results from implementing a load capacitor. However, we note that increasing the number of windings to the optimal $N$ may challenging in terms of practical spacing in the presented devices. In addition, the present model assumes ideal coil windings and does not account for manufacturing related factors such as non-uniform windings or imperfect flux linkage between the coils and the magnetic components, which may reduce the achievable power output in practical devices. Note, however, that in terms of maximum power output, it is the amount of copper that matters and not the specific combination of winding turns and wire radius. Thus, it may be more practical to increase the wire radius, as done by Bahl et al. \cite{Bahl2024DesignHarvester}, and use fewer turns when optimizing the coil design.

\section{Discussions and future perspectives}
For obtaining the maximum power from a TMG device, the load impedance should match the internal impedance of the syxstem.  
Matching the internal resistance of the coils is relatively straightforward with a varying load resistor. However, the reactance of the system is given by the self-inductance $L$, which depends on the permeability of the MCM beds changing over time. Our analysis demonstrates that the self-inductance required for the purpose of impedance matching is given by Eq. \eqref{eq:selfLapprox}, which is entirely determined by the average reluctance of the MCM beds and not, as one might expect, by the total reluctance of the magnetic circuit. In a real device, a varying load capacitor could potentially be used to match the instantaneous reactance of the coils, and thus optimize power further as was shown in section 3.1.

It is important to stress that the presented model could also be implemented to optimize other components of the TMG. From Eq. \eqref{Power_optimized_reluctance}, which describes the maximized TMG power, it is clear that several parameters affect the power, including $B_{\text{rem}}$, $V_M$, $a$, $\mu_M$ and $\overline{\mc{R}}_{\text{bed}}$. For example, $B_{\text{rem}}$ and $V_M$ are directly linked to the permanent magnet and its physical properties. As an example, a future study could numerically optimize these parameters e.g. by considering realistic values of $B_{\text{rem}}$ and $V_M$ corresponding to commercially available permanent magnets. Such a study could ideally be complemented by an experimental validation, using the same permanent magnets and geometries. Additionally, with minor modifications, the presented model could also be adapted to optimize other key components, such as the size, shape and configuration of the thermomagnetic material.  

Regarding future improvements to the presented model, as discussed in the validation section, the model does not yet include common loss mechanisms such as magnetic flux leakage and eddy currents. Future work should include these mechanisms as well as the extension to three dimensions to accurately capture the flux losses. Here, we note that once the magnetic fields have been computed by finite element methods, they could, at least in an approximate way, be included in the magnetic equations of our coupled lumped model to estimate the effect of magnetic leakage. 

Additionally, the estimation of the two model material parameter, $\overline{\mu}$ and $\Delta_{\mu}$, can be further refined. In future work, this could be achieved by implementing a heat transfer model that takes the material characteristics and reservoir temperatures as input, such as \cite{deJesus2024ThermodynamicMaterials,Correa2023ThermodynamicMotor,Almanza2017NumericalCycle} to enable calculation of $\overline{\mu}$ and $\Delta_{\mu}$ as output. In general, a coupling between the findings of these models with the one presented in this work is expected to provide new insights into optimal operating conditions, for example the operating frequency and the temperature difference between the heat reservoirs. Moreover, integrating this approach in the design phase would allow more accurate optimization of future TMG prototypes and their Carnot efficiency.

\section*{Data statement}
All data shown and used in this work are available at the data repository at Ref.\cite{bjørk_nilsson_insinga_de_angelis_2025}.

\section{Conclusion}
In this work, we presented a study on the influence of coils on the performance of a thermomagnetic generator (TMG) using an analytical and numerical model that explicitly couples its magnetic and electric circuits. The governing ordinary differential equations describing the dynamics of the coupled TMG system were derived, and the analytical model showed that power had a linear dependence on the coil volume. Further, it was shown that this dependence is independent of the specific combination of wire radius and coil turns, providing new insights for the design processes of future TMG coils. 

Using the developed model, we analytically determined values for the optimal load resistance, load reactance, and magnet reluctance that maximize the power produced by the TMG. We showed that the load impedance that maximize power, required the implementation of a load capacitor matching the two coil's self-inductance. Interestingly, the derived expression for the self-inductance in the electric circuit, Eq. \eqref{eq:selfLapprox}, required it to be inversely proportional to the average reluctance of the MCM bed, which aligned it with the magnetic circuit. Additionally, to maximize power, we showed that the optimal reluctance of the permanent magnet had to be half of the average reluctance of the MCM beds, Eq. \eqref{optimal_reluctance}.

The available energy of a TMG or the rate of energy dissipated in the coils through Joule heating, were shown to be directly related to the magnetic hysteresis of the MCM beds. A piecewise linear model was implemented in the numerical model to account for magnetic saturation of the MCM, which were shown to realistically restrict the available energy of a TMG. The model was validated with experimental data from literature, and finally used to study prototype TMGs presented in literature. This evaluation have shown that the power of the devices could be increased by a factor of 10-400 times, had larger coils been used in the prototypes.

\section*{Acknowledgements}
\noindent This work was funded by the Independent Research Fund Denmark, grant 3164-00057B on Magnetothermal harvesting of low-grade waste heat.

\bibliographystyle{plain}

\end{document}